\documentclass[smallextended]{svjour3}       
\smartqed  
\usepackage{graphicx}
\usepackage{bm}
\DeclareGraphicsExtensions{.png,.pdf,.jpg}
\usepackage[hmargin=1in,vmargin=1in]{geometry}
\usepackage{verbatim}
\usepackage{amssymb, amsmath}
\usepackage{booktabs}
\usepackage{hyperref}
\usepackage{appendix}
\usepackage{tcolorbox}
\usepackage{longtable}
\usepackage{afterpage}
\usepackage{pdflscape}
\usepackage{capt-of}
\usepackage{lscape}
\usepackage{subcaption}
\usepackage{url}
\usepackage{multirow}
\usepackage{wrapfig}
\usepackage{lscape}
\usepackage{rotating}
\usepackage{epstopdf}
\usepackage{float}
\usepackage{lineno,hyperref}
\modulolinenumbers[5]
\DeclareMathOperator{\Tr}{Tr}
\newcommand{\Polya}{P\'{o}lya}
\DeclareMathOperator*{\argmin}{arg\,min}

\journalname{Neuroinformatics}

\begin{document}

\title{{\it BVAR-Connect}: A Variational Bayes Approach to Multi-Subject Vector Autoregressive Models for Inference on Brain Connectivity Networks}

\titlerunning{{\it BVAR-Connect:} Bayesian VAR Models for Brain Connectivity}        

\author{Jeong Hwan Kook \and Kelly A. Vaughn \and Dana M. DeMaster \and Linda Ewing-Cobbs \and Marina Vannucci
}

\institute{Jeong Hwan Kook and Marina Vannucci \at Department of Statistics\\ Rice University\\ Houston, TX 77005, USA\\
\email{eric.jeong.kook@gmail.com, marina@rice.edu}
\and Kelly A. Vaughn, Dana M. DeMaster and Linda Ewing-Cobbs \at Dept. of Pediatrics\\ Children's Learning Institute\\ University of Texas Health Science Center\\ Houston, Texas 77030
}

\date{{\it Neuroinformatics} (2020) Software Original Article; Accepted 29 April, 2020}

\maketitle
	
\begin{abstract}
In this paper we propose {\it BVAR-connect}, a variational inference approach to a Bayesian multi-subject vector autoregressive (VAR) model for inference on effective brain connectivity based on resting-state functional MRI data. The modeling framework uses a Bayesian variable selection approach that flexibly integrates multi-modal data, in particular structural diffusion tensor imaging (DTI) data, into the prior construction. The variational inference approach we develop allows scalability of the methods and results in the ability to estimate subject- and group-level brain connectivity networks over whole-brain parcellations of the data.  We provide a brief description of a user-friendly MATLAB GUI released for public use. We assess performance on simulated data, where we show that the proposed inference method can achieve comparable accuracy to the sampling-based Markov Chain Monte Carlo approach but at a much lower computational cost.  We also address the case of subject groups with imbalanced sample sizes.  Finally, we illustrate the methods on resting-state functional MRI and structural DTI data on children with a history of traumatic injury. 
\end{abstract}

\section{Introduction}\label{sec:intro}
In recent years, advances in neuroimaging techniques have led to various methods of modeling brain connectivities. To this day, functional magnetic resonance imaging (fMRI) is still one of the most popular techniques for measuring and mapping brain activity, mostly due to its noninvasive nature. This technique measures blood oxygenation level dependent (BOLD) contrast, i.e. the difference in magnetization between oxygenated and deoxygenated blood arising from changes in regional cerebral blood flow. Changes in BOLD response are treated as a proxy for changes in neurological activity.  
	
In this paper, we are interested in studying the influence that one neural system exerts over another one based on resting-state fMRI data, for the estimation of brain connectivity networks. 
This type of connectivity is commonly referred to as effective connectivity. Statistical approaches to modeling effective connectivity among interconnected brain regions of interest include dynamic causal modeling (DCM) \cite{friston2003dynamic}, structural equational modeling (SEM) \cite{mclntosh1994structural}, Bayesian networks (BNs) \cite{li2008,rajapakse2007learning} and Granger causality modeling via vector autoregressive (VAR) models \cite{granger1969,roebroeck2005}. DCM and SEM aim to infer effective connectivity for pre-specified connectivity patterns, and are both best used as confirmatory techniques to test pre-defined hypotheses about neural activity \cite{friston2011functional}. Bayesian networks estimate effective connectivity by modeling brain networks as directed acyclic graphs, therefore ignoring the high prevalence of reciprocal connections that commonly renders brain connectivity cyclic \cite{friston2011functional}. Granger causality is based on the simple notion that causes both precede and help predict their effects. This approach does not require pre-specification of any connectivity patterns but rather infers effective connectivity by estimating and comparing coefficients from two VAR models.  It is important to notice that, even though such methods allow inference on directed
connections between brain regions, causality between fMRI signals does not translate into causality of the corresponding neuronal activity \cite{Wen2013}. In spite of this, fMRI connectivity studies have been proven helpful in understanding the role that connectivity patterns, and their disruption, play in mental health disorders and brain diseases. 
	
Several researchers have proposed approaches to VAR models for single and multiple subjects data \cite{deshpande2009multivariate,gorrostieta2012investigating,gorrostieta2013hierarchical,yu16,Chiang17}. However, one limitation of these approaches is that they can only handle a small number of brain regions. This has forced investigators to focus on selected brain regions, perhaps within specific brain networks,  or to resort to dimension reduction via ICA \cite{calhoun2001method}. Here we focus our attention on the Bayesian multi-subject VAR modeling approach developed in \cite{Chiang17}. This modeling framework uses a Bayesian variable selection approach to allow for simultaneous inference on effective connectivity at both the subject- and group-level. We exploit the flexibility of the approach to integrate multi-modal data, in particular structural DTI data, into the prior construction, thus encouraging connectivity between structurally connected regions. Instead of the probit-like regression construction used by \cite{Chiang17}, we employ a logit prior and the data augmentation method proposed by \cite{Polson13}, that allows for efficient posterior updates of some of the model parameters.  For posterior inference, we address issues of scalability of the models by developing a novel variational inference approach that avoids the computational bottlenecks of the sampling-based Markov chain Monte Carlo approach used by \cite{Chiang17}. As the number of brain regions increases, the computation time needed for the MCMC samples to converge increases dramatically.  Variational inference, on the other hand, turns posterior inference into an optimization problem, which aims at finding a class of approximating distributions that minimizes the Kullback-Leibler divergence to the exact posterior distribution. This allows scalability of the models and results in the ability to estimate brain connectivity networks over whole-brain parcellations of the data, a type of analysis which is not feasible under the MCMC framework of \cite{Chiang17}.
 
We name our proposed method {\it BVAR-connect} and provide a brief description of a user-friendly MATLAB GUI released for public use. We show through simulations that,  with respect to sampling-based MCMC approaches, the variational inference method results in a dramatic speedup of model convergence without compromising the accuracy of the estimation. We also address the case of subject groups with imbalanced sample sizes. 
 
We illustrate performance of our method via an application to resting-state fMRI and structural DTI data collected on children with a history of traumatic injury.  Consistent with structural findings, our results suggest weaker functional connectivity across regions in children with traumatic brain injury (TBI) relative to those with extracranial injury (EI) to body regions excluding the brain. They also show an increased number of connections in children with TBI, indicating that the brain may be using compensatory functional connections in response to damage to other functional and/or structural connections. 
	
The rest of the paper is organized as follows.  In Section \ref{sec:methods}, we review the model and the prior construction and then introduce the proposed variational approach. We also briefly describe the Matlab GUI {\it BVAR-connect}.  In Section \ref{sec:simulation}, we assess performance of our proposed method against the MCMC approach of \cite{Chiang17} using simulated data. In Section \ref{sec:fMRI} we illustrate our method on resting-state functional MRI and structural DTI data collected on children with a history of traumatic injury. Section \ref{sec:discussion} concludes the paper with a discussion on limitations and future directions.
	
\section{Materials and Methods}
\label{sec:methods}
	
\subsection{Bayesian vector autoregressive models}
\label{sec:likelihood}
Let $x_{t,j}^{(s)}$ be the BOLD fMRI response of subject $s$ at time $t$ in region $j$, for $t=1,\ldots,T, j=1,\ldots,R$ and $s=1, \ldots,n$. Here, regions can be viewed as portions of the brain obtained via brain parcellations derived from anatomical landmarks or Independent Components Analysis (ICA) \cite{calhoun2001method}. We assume that the $n$ subjects belong to $G$ distinct groups and that each group exhibits different brain activities. We let $\eta_{s}$ be the known group label for subject $s$, with $\eta_{s}=g$ if subject $s$ belongs to group $g\in G$. Temporal correlation is modeled through a multivariate linear vector autoregressive (VAR) process of order $L$ as

\begin{equation}\label{eqn:1}
\left(x_{t}^{(s)}\mid\eta_{s}=g,\phi_{l,g}^{(s)},\Xi\right)=\sum_{l=1}^{L}\phi_{l,g}^{(s)}x_{t-l}^{(s)}+e_{t}^{(s)},
\end{equation}
with $e_{t}^{(s)}\sim N(0,\Xi)$ and where $x_{t}^{(s)}=[x_{t,1}^{(s)},\ldots,x_{t,R}^{(s)}]'$ is the $R\times1$ vector of fMRI BOLD responses at time $t$ for subject $s$ and where the parameters $\phi_{l,g}^{(s)}$, for $l,g=1,\ldots,R$ represent the lag-specific effective connectivities between the $R$ regions for subject $s$. We assume $\Xi=diag(\zeta_{1},\ldots\zeta_{R})$ and place inverse gamma priors on the diagonal elements $\zeta_{j}\sim IG(h_{1},h_{2})$ for,$j=1,\ldots,R$. Prior to analysis, we assume that the data have been centered.  
	
The VAR model shown in \eqref{eqn:1} can be written in a multivariate regression form as
\begin{equation}
\underbrace{x_{t}^{'(s)}}_{1\times R}=\underbrace{u_{t}^{'(s)}}_{1\times RL}\underbrace{B_{t}^{(s)}}_{RL\times R}+\underbrace{e_{t}^{'(s)}}_{1\times R},\nonumber
\end{equation}
where $u_{t}^{'(s)}=[x_{t-1}^{'(s)},x_{t-2}^{'(s)},\ldots,x_{t-L}^{'(s)}]$ is the $1\times RL$ vector of concatenated lagged BOLD data, and $B_{g}^{(s)}=[\phi_{1,g}^{(s)},\phi_{2,g}^{(s)},\ldots,\phi_{L,g}^{(s)}]'$ the $RL\times R$ matrix of concatenated subject-specific effective connectivities. For $T$ time points, we have
\begin{equation}
\underbrace{X^{(s)}}_{(T-L)\times R}=\underbrace{U^{(s)}}_{(T-L)\times RL}\underbrace{B_{g}^{(s)}}_{(RL)\times R}+\underbrace{E^{(s)}}_{(T-L)\times R}. \nonumber
\end{equation}
Using the $vec$ operator, which converts a matrix into a column vector by stacking each column on top of each other, we let $\underline{x}^{(s)}=vec(X^{(s)})$, $\underline{B}_{g}^{(s)}	=vec(B_{g}^{(s)}) $ and $\underline{e}^{(s)}	=vec(E^{(s)})$.  We can then rewrite the Bayesian VAR model as
\begin{equation}
\label{eqn:2}
\underline{x}^{(s)}=\left(I\otimes U^{(s)}\right)\underline{B}_{g}^{(s)}+\underline{e}_{g}^{(s)},
\end{equation}
with $\underline{e}^{(s)}\sim N(0,\Xi\otimes I)$ and where $\otimes$ denotes the Kronecker product. We then have that, for subject $s$ with group label $\eta_{s}=g$, $\underline{x}^{(s)}$ is normally distributed as 
\begin{equation}\label{eqn:3}
\left(\underline{x}^{(s)}\mid\eta_{s}=g,\underline{B}_{g}^{(s)},\Xi\right)\sim N\left((I\otimes U^{(s)})\underline{B}_{g}^{(s)},\Xi\otimes I\right),
\end{equation}
with parameters in $\underline{B}_{g}^{(s)}$ capturing the subject-level effective connectivities.
	
\subsection{Prior construction}
We adopt the prior construction of \cite{Chiang17} and model the subject-level parameters $\underline{B}_{g}^{(s)}$ in \eqref{eqn:3} as random deviations from a baseline process following a normal distribution. Specifically, for all subjects $s$ such that $\eta_{s}=g$, with $g=1,\ldots, G$, we have	
\begin{equation}\label{eqn:4}
p\left(\underline{B}_{g}^{(s)}\mid\Omega^{(g)},\Sigma^{(g)}\right)=N\left(\Omega^{(g)},\Sigma^{(g)}\right),
\end{equation}
with baseline process $\Omega^{(g)}$ and a $LR^{2}-by-LR^{2}$ diagonal matrix $\Sigma^{(g)}=diag(\sigma_{1}^{(g)},\ldots\sigma_{LR^{2}}^{(g)})$. Next, spike-and-slab mixture priors are employed to select non-zero effective connectivities across and within lags at the group level. We use discrete constructions that employ a spike distribution at zero \cite{GeorgeMcCulloch97,brown}. More specifically, for every $\omega_{k}^{(g)}\in\Omega^{(g)}, k = 1,\ldots, LR^{2}$, a latent binary indicator $\gamma_{k}^{(g)}$ is introduced to indicate a non-zero effective connectivity, i.e.,  $\gamma_{k}^{(g)}=1$ if $\omega_{k}^{(g)}\neq 0$, and zero otherwise. To account for spatio-temporal smoothness, an intrinsic conditional autoregressive (ICAR) distribution \cite{banerjee} is used as the slab portion of the mixture prior as
\begin{equation}\label{eqn:5}
\omega_{k}^{(g)}\sim\gamma_{k}^{(g)}N\left(\frac{\sum_{k'=1}^{LR^{2}}S_{kk'}\omega_{k'}^{(g)}}{\sum_{k'=1}^{LR^{2}}S_{kk'}},\frac{q}{\sum_{k'=1}^{R^{2}}S_{kk'}}\right)+(1-\gamma_{k}^{(g)})\delta_{0}(\omega_{k}^{(g)}),
\end{equation}
where $\delta_{0}(\omega_{k}^{(g)})$ is the spike at zero, $S$ is a $LR^{2}\times LR^{2}$ symmetric matrix of binary elements that controls smoothness, and $q$ is a variance tuning parameter. The ICAR prior allows for smoothness by encouraging effective connectivities to vary smoothly across temporal lags or groupings of VAR coefficients. The specification $S = I$ corresponds to the case
of no prior knowledge about spatial and temporal smoothness. Similarly, for $\gamma_{k}^{(g)}=1$ we place $\sigma_{k}^{(g)}=\xi_{1}^{(g)}\sim IG(a_{1}^{(g)},b_{1}^{(g)})$ and $\sigma_{k}^{(g)}=\xi_{0}^{(g)}\sim IG(a_{0}^{(g)},b_{0}^{(g)})$ for all $k$ such that $\gamma_{k}^{(g)}=0$.
	
Prior construction \eqref{eqn:5} is extremely flexible and allows the incorporation of external information, for example in the form of prior knowledge or additional data, on the presence or absence of connectivities. Here we incorporate such information, when available, by specifying a logit regression prior on $\gamma_{k}^{(g)}$ as
\begin{equation}\label{eqn:6}
p(\gamma_{k}^{(g)})=\frac{\exp(\alpha_{0}^{(g)}+\alpha_{1}^{(g)}N_{k}^{(g)})^{\gamma_{k}^{(g)}}}{1+\exp(\alpha_{0}^{(g)}+\alpha_{1}^{(g)}N_{k}^{(g)})}, 
\end{equation}
where $N_{k}^{(g)}$ is the external available measure of connectivity between the two regions indexed by the $k$th element of $\Omega^{(g)}$. This formulation is somewhat different from the construction of \cite{Chiang17}, who used a probit-like regression prior. Our prior formulation, in particular, allows for efficient posterior inference via data augmentation, see next Section. Furthermore, in the application of this paper, we use this construction to incorporate structural information captured by Diffusion Tensor Imaging data, as it is known that the presence of white matter connection between brain regions generally leads to an increased chance of functional or effective connectivity \cite{greicius2009resting,deco2011,kang17,higgins18}. Our prior construction allows the strength  of a structural connection to inform the prior probability of having a non-zero effective connectivity. The parameter $a^{(g)}_{0}$ regulates the prior probability of non-zero effective connectivity when structural connectivity is not present (i.e., $N_k^{(g)}=0$). Finally, $a^{(g)}_{1}$ is modeled using a normal prior as $a^{(g)}_{1}\sim N(\omega^{(g)},\tau^{2(g)}),$ for $g=1,\ldots,G$.  In cases where external information is not available, the logit prior in \eqref{eqn:6} can be replaced with a Beta-Bernoulli conjugate prior
\begin{equation}\label{eqn:beta-bernoulli}
p(\gamma_{k}^{(g)})=\pi^{(g)\gamma_{k}^{(g)}}(1-\pi^{(g)})^{1-\gamma_{k}^{(g)}}, 
~~~\pi^{(g)}\sim Beta(e^{(g)},f^{(g)}),
\end{equation}	
with $e^{(g)}$ and $f^{(g)}$ hyper-parameters regulating the prior probability of non-zero connectivity.

\subsection{Model fitting using variational Bayes}\label{sec:VB}
For posterior inference, we propose a novel approach that uses variational inference \cite{Bishop2000,beal2003,Blei16}.  Variational approaches turn inference into an optimization problem, making posterior inference scalable and computationally faster than sampling-based Markov chain Monte Carlo methods. Typically, variational approaches provide good estimates of means; however, they tend to underestimate posterior variances and the correlation structure of the data \cite{bishop2006}. This shortcoming can still be an acceptable trade-off for our inferential purposes, as we are mainly interested in the identification of non-zero edges in networks. Indeed, VB approaches have been effectively used in Bayesian models for the analysis of fMRI data \cite{Penny2003,Flandin2007,Woolrich2004,zhang2016,kook19}. 
	
The key idea of VB methods is to find an approximation to the posterior distribution that minimizes the Kullback Leibler (KL) divergence.  In the context of our model, given the model parameters and latent variables $z=\{\underline{B}_{g}^{(s)},\Omega^{(g)},$ $\xi_{1}^{(g)},\xi_{0}^{(g)},\alpha_{1}^{(g)},\zeta_{j},\gamma_{k}^{(g)}\}$, for $j=1,\ldots,R$, and $k=1,\ldots,LR^{2}$, the approach specifies a family $\mathcal{D}$ of densities $q_{\phi}(z)\in\mathcal{D}$, with free variational parameters $\phi$,  as a candidate approximation to the target posterior distribution $p(z\mid x)$ and then chooses $\phi$ to minimize
\begin{equation}\label{eqn:7}
\begin{split}
q_{\phi^* } &=\argmin_{q_{\phi}(z)\in\mathcal{D}} KL(q(z)\rVert p(z\mid x))  
=\log{p(x)} - \Big(\mathbb{E}_{q_{\phi}(z)}[\log p(z,x)]+\mathbb{H}[q_{\phi}(z)]\Big),
\end{split}
\end{equation}
with $\mathbb{H}[q_{\phi}(z)]$ the entropy of $q_{\phi}(z)$. This amounts to maximize the so called evidence lower bound
\begin{equation}\label{eqn:8}
ELBO(q)=\mathbb{E}_{q_{\phi}(z)}[\log p(z,x)]+\mathbb{H}[q_{\phi}(z)].
\end{equation}
Clearly, the complexity of the approximating class $q(z)$ determines the complexity of the optimization procedure.  Having a more expressive variational family potentially improves the fidelity of the approximation, but in practice it entails deriving non-trivial, model-specific derivations for the optimization problem.  Mean-field approximation schemes, which assume that the latent variables are mutually independent and each governed by a distinct factor in the variational density, are commonly employed. Here, we implement a scheme that takes advantage of data augmentation, to obtain an efficient sampling algorithm. 
	
\subsubsection{Polya-gamma augmentation}
Given our logistic prior construction \eqref{eqn:6}, we can use the \Polya-Gamma augmentation scheme proposed by \cite{Polson13} to implement efficient closed-form VB updates that exploit the conditional conjugacy of the latent parameters.  A random variable $\omega$ following a \Polya-Gamma distribution with parameters $b>0, c\in \mathcal{R}$ is defined as
\begin{equation}\label{eqn:11}
\omega\overset{D}{=}\frac{1}{2\pi^{2}}\sum_{k=1}^{\infty}\frac{g_{k}}{(k-1/2)^{2}+c^{2}/(4\pi^{2})}, 
\end{equation}
with $\mathbb{E}[\omega] =\frac{b}{2c}\tanh(c/2)$ and where the $g_{k}\sim Ga(b,1)$ are independent gamma random variables and $\overset{D}{=}$ indicates equality in distribution. The main result from \cite{Polson13} is that given a random variable $\omega$ with density $p(\omega)\sim PG(b,0),b>0,$ the following integral identity holds for all $a\in\mathcal{R}$:
\begin{equation}\label{eqn:12}
\begin{split}
\frac{(e^{\psi})^{a}}{(1+e^{\psi})^{b}}=2^{-b}e^{\kappa\psi}\int^{\infty}_{0}e^{-\omega\psi^{2}/2}p(\omega)d\omega
=2^{-b}e^{\kappa\psi}\mathbb{E}_{\omega}[\exp(-\omega\psi^{2}/2)], 
\end{split}
\end{equation}
where $\kappa=a-b/2$. Additionally, the conditional distribution $p(\omega\mid\psi)$ arising from the above integrand is also in the \Polya-Gamma class $p(\omega\mid\psi)\sim PG(b,\psi)$. For more details regarding the derivation of the result, we refer interested readers to \cite{Polson13}. Here,
we apply identity \eqref{eqn:12} to the logistic prior \eqref{eqn:6}. Let $\phi_{k}^{(g)}$ be a \Polya-Gamma distributed random variable for $k=1,\ldots,LR^{2},$ and $g=1,\ldots, G$. It follows that
\begin{equation}\label{eqn:13}
\begin{split}
p(\gamma_{k}^{(g)})&=\frac{\exp(\alpha_{0}^{(g)}+\alpha_{1}^{(g)}N_{k}^{(g)})^{\gamma_{k}^{(g)}}}{1+\exp(\alpha_{0}^{(g)}+\alpha_{1}^{(g)}N_{k}^{(g)})} \\ 
&\propto \exp\left((\gamma_{k}^{(g)}-\frac{1}{2})[\alpha_{0}^{(g)}+\alpha_{1}^{(g)}N_{k}^{(g)}]\right)\mathbb{E}_{\phi_{k}^{(g)}}\left[\exp(-\phi_{k}^{(g)}(\alpha_{0}^{(g)}+\alpha_{1}^{(g)}N_{k}^{(g)})^{2}/2)\right].
\end{split}
\end{equation}
Note that the expectation terms on the right hand side of the equation is the kernel of a Gaussian likelihood in $\alpha_{1}^{(g)}$.  Consequently, the conditional posterior of $\phi_{k}^{(g)}$ is \Polya-Gamma distributed with parameters $(1,\alpha_{0}^{(g)}+\mathbb{E}[\alpha_{1}^{(g)}]N_{k}^{(g)})$ and expectation $\mathbb{E}[\phi_{k}^{(g)}]=\frac{\tanh\left((\alpha_{0}^{(g)}+\mathbb{E}[\alpha_{1}^{(g)}]N_{k}^{(g)})/2\right)}{2(\alpha_{0}^{(g)}+\mathbb{E}[\alpha_{1}^{(g)}]N_{k}^{(g)})}$. By placing a normal variational distribution for $\alpha_{1}^{(g)}$  as $q(\alpha_{1}^{(g)})=N(\mu_{\alpha_{1}^{(g)}},\sigma_{\alpha_{1}^{(g)}}^{2})$, we can exploit conjugacy of the likelihood and have a closed-form update rule for the parameter.
	
\subsubsection{Mean-field factorized distribution}
The variational distributions for the remaining set of parameters are defined as follows: 
\begin{equation}
\label{eqn:14}
\begin{split}
q(\xi_{1}^{(g)};c_{1}^{(g)},d_{1}^{(g)})&\sim IG(c_{1}^{(g)},d_{1}^{(g)})\\
q(\xi_{0}^{(g)};c_{0}^{(g)},d_{0}^{(g)})&\sim IG(c_{0}^{(g)},d_{0}^{(g)})\\
q(\underline{\beta}_{g}^{(s)}\mid U_{g}^{(s)},\Sigma_{g}^{(s)})&\sim MVN(U_{g}^{(s)},\Sigma_{g}^{(s)})\\
q(\zeta_{j}\mid z_{1j},z_{2j})&\sim IG(z_{1j},z_{2j}),\\
q(\tilde{\omega}_{k}^{(g)})&\sim N(\mu_{k}^{(g)},s_{k}^{2(g)})\\
q(\gamma_{k}^{(g)})&\sim Bernoulli(\nu_{k}^{(g)}),
\end{split}
\end{equation}
for $g=1,\ldots,G,\, s = 1,\ldots,n,\, k= 1,\ldots, LR^{2},$ and $j=1,\ldots, R$.  Here, we follow the formulation of \cite{Titsias2011} for spike-and-slab priors and introduce a new auxiliary variable $\tilde{\omega}_{k}^{(g)}$ such that $\omega_{k}^{(g)}=\gamma_{k}^{(g)}\tilde{\omega}_{k}^{(g)}$ for $k=1,\ldots, LR^{2},$ and $g=1,\ldots, G$. This representation allows us to model $\{\tilde{\omega}_{k}^{(g)},\gamma_{k}^{(g)}\}$ jointly as 
\begin{equation}
\label{eqn:9}	
q\left(\tilde{\omega}_{k}^{(g)}\mid\gamma_{k}^{(g)}\right)=\begin{cases}
N\left(\tilde{\omega}_{k}^{(g)}\mid\mu_{k}^{(g)},s_{k}^{2(g)}\right) & \text{if }\gamma_{k}=1\\
N\left(\tilde{\omega}_{k}^{(g)}\mid\frac{\sum_{k'=1}^{LR^{2}}S_{kk'}\omega_{k'}^{(g)}}{\sum_{k'=1}^{LR^{2}}S_{kk'}},\frac{q}{\sum_{k'=1}^{LR^{2}}S_{kk'}}\right) & \text{\text{if \ensuremath{\gamma_{k}}=0 }}
\end{cases}
\end{equation}
and $q(\gamma_{k}^{(g)})=\nu_{k}^{(g)\gamma_{k}^{(g)}}(1-\nu_{k}^{(g)})^{1-\gamma_{k}^{(g)}}$,
where $\{\mu_{k}^{(g)},s_{k}^{2(g)},\nu_{k}^{(g)}\}$ are the variational parameters. The proposed representation yields a marginal $q(\tilde{\omega}_{k}^{(g)})$ which has $2^{LR^{2}}$ components and returns a posterior distribution equal to the  prior distribution when $\gamma_{k}^{(g)}=0$. In sum, the approximating distribution $Q$ can be written as
\begin{equation}
\label{eqn:15}
\begin{split}
Q\left(\Theta\right)=\prod_{s=1}^{n}q\left(\underline{\beta}_{g}^{(s)}\right)\prod_{j=1}^{R}q\left(\zeta_{j}\right)\prod_{g=1}^{G}q\left(\alpha_{1}^{(g)}\right)q\left(\xi_{1}^{(g)}\right)q\left(\xi_{0}^{(g)}\right) \prod_{k=1}^{LR^{2}}q\left(\tilde{\omega}_{k}^{(g)}\mid\gamma_{k}^{(g)}\right)q\left(\gamma_{k}^{(g)}\right)q\left(\phi_{k}^{(g)}\right),
\end{split}
\end{equation}
where $\Theta$ denotes the variational parameters to be optimized. In cases where the model is fitted with the Beta-Bernoulli prior \eqref{eqn:beta-bernoulli}  instead of the logit construction \eqref{eqn:6}, the variational distributions for $\alpha_{1}^{(g)}$ and $\phi_{k}^{(g)}$, introduced via the \Polya-Gamma construction, are omitted from the model and replaced by a variational distribution for $\pi^{(g)}$ in \eqref{eqn:beta-bernoulli} using a beta distribution as $q(\pi^{(g)})\sim Beta(m^{(g)},n^{(g)})$.
	
\subsubsection{Variational Bayes algorithm}
\label{sec:VB Algorithm}
A generic VB algorithm for posterior inference comprises updating the variational parameters to minimize the ELBO.  A common algorithm is coordinate ascent variational inference (CAVI) \cite{bishop2006}. For any latent parameter of interest $z_{j}$, the ELBO can be maximized by performing a minimization over $q(z_{j})$, which leads to the following equation
\begin{equation}
\label{eqn:16}
q^{*}_{j}(z_{j})\propto  \exp \{\mathbb{E}_{j}[\text{log }p(z_{j},z_{-j},X)]\}, 
\end{equation}
where the expectation is taken with respect to the currently fixed variational density over $\prod_{i\neq j}q_{i}(z_{i}))$. For conjugate models, the right hand side of \eqref{eqn:16} has the same functional form as the prior and the expression represents an implicit solution for the variational posterior. 
	
Given the factorization of $Q$ in \eqref{eqn:15},  the posterior $q(\tilde{\omega}_{k}^{(g)},\gamma_{k}^{(g)})$ can be computed as
\begin{equation}
\label{eqn:17}
\begin{split}
q\left(\tilde{\omega}_{k}^{(g)},\gamma_{k}^{(g)}\right)=\frac{1}{\mathcal{Z}}\prod_{s:\eta_{g}=g}\exp\left\{\mathbb{E}\left[\log p(\underline{B}_{g}^{(s)}\mid\Omega^{(g)},\Sigma^{(g)})\right] \right\}
N\left(\tilde{\omega}_{k}^{(g)}\mid 0,\sigma_{w}^{2}\right)\nu_{k}^{(g)\gamma_{k}^{(g)}}\left(1-\nu_{k}^{(g)}\right)^{1-\gamma_{k}^{(g)}},
\end{split}
\end{equation}
where $\mathcal{Z}$ is the partition function, and the expectation is taken with respect to $Q(\Theta)$ with $(\tilde{\omega}_{k}^{(g)},\gamma_{k}^{(g)})$ removed. To get an explicit expression of the above posterior, we first find the marginal distribution $q(\gamma_{k}^{(g)}=1)$ by setting $\gamma_{k}^{(g)}=1$ in \eqref{eqn:17} and integrating out $\tilde{\omega}_{k}^{(g)}$, such as $q(\gamma_{k}^{(g)}=1) =\int q(\tilde{\omega}_{k}^{(g)},1)d\tilde{\omega}_{k}^{(g)}$, and then repeat the same procedure for $\gamma_{k}^{(g)}=0$.  It is easy to show that
\begin{equation}
\label{eqn:19}
\nu_{k}^{(g)}=q(\gamma_{k}^{(g)}=1)=\frac{1}{1+\exp(-\rho_{k}^{(g)})},
\end{equation}
where $\rho_{k}^{(g)}=\log q(\gamma_{k}^{(g)}=1) - \log q(\gamma_{k}^{(g)}=0)$. We can then write \eqref{eqn:17} as the product of a conditional and a marginal
\begin{equation}
\label{eqn:20}
\begin{split}
q\left(\tilde{\omega}_{k}^{(g)},\gamma_{k}^{(g)}\right)&=
q\left(\tilde{\omega}_{k}^{(g)}\mid\gamma_{k}^{(g)}\right)q\left(\gamma_{k}^{(g)}\right)=\\
&=N\left(\tilde{\omega}_{k}^{(g)}\mid\gamma_{k}^{(g)}\mu_{k}^{(g)},\gamma_{k}^{(g)}s_{k}^{2(g)}+\left(1-\gamma_{k}^{(g)}\right)\frac{q}{\sum_{k'=1}^{LR^{2}}S_{kk'}}\right)\nu_{k}^{(g)\gamma_{k}^{(g)}}\left(1-\nu_{k}^{(g)}\right)^{1-\gamma_{k}^{(g)}}.
\end{split}
\end{equation}
	
After initializing the variational parameters $\Theta$, the algorithm consists of repeating the following steps to update $\Theta$ until convergence of the ELBO is met:
\begin{enumerate}
\item \textbf{Update $\{U_{g}^{(s)},\Sigma_{g}^{(s)}\}$ of $\underline{\beta}_{g}^{(s)}$, for $s=1,\ldots, n$}: We exploit conjugacy of the model and the approximating variational distribution, arriving to a closed-form expression for $\{U_{g}^{(s)},\Sigma_{g}^{(s)}\}$. The update is performed one subject at a time.
\item \textbf{Update $\{z_{1j},z_{2j}\}$ of $\zeta_{j}$, for $j=1,\ldots, R$}: 
By conjugacy, closed form expressions can be obtained for $\{z_{1j},z_{2j}\}$. The update is performed in a vectorized manner, such that the parameters are updated jointly for all $j=1,\ldots, R$.
\item \textbf{Update $\{c_{1}^{(g)},d_{1}^{(g)}\},\{c_{0}^{(g)},d_{0}^{(g)}\}$ of $\xi_{1}^{(g)},\xi_{0}^{(g)}$ for $g=1,\ldots G$}:
By exploiting conjugacy, the pairs $\{c_{1}^{(g)},d_{1}^{(g)}\},\{c_{0}^{(g)},d_{0}^{(g)}\}$ are updated one group at a time.		
\item \textbf{Update $\{\mu_{k}^{(g)},s_{k}^{2(g)}\}$ of $\tilde{\omega}_{k}^{(g)}$ and $\nu_{k}^{(g)}$ of $\gamma_{k}^{(g)}$ for $k=1,\ldots LR^{2}, g=1,\ldots, G$}:
For each group $g$, we first update $\{\mu_{k}^{(g)},s_{k}^{2(g)}\}$ and then approximate $\nu_{k}^{(g)}$ using Monte Carlo samples. The ordering of $k$ is determined by a random permutation.
\item \textbf{Update $\{\mu_{\alpha_{1}^{(g)}},\sigma_{\alpha_{1}^{(g)}}^{2}\}$ of $\alpha_{1}^{(g)}$, for $g=1,\ldots G$}: We follow the data augmentation procedure described in \eqref{eqn:13} to obtain closed form updates for $\{\mu_{\alpha_{1}^{(g)}},\sigma_{\alpha_{1}^{(g)}}^{2}\}$. Update is performed one group at a time.
\item \textbf{Update $\phi_{k}^{(g)}$ for $k=1,\ldots, LR^{2}, g=1,\ldots, G$}: We follow \cite{Polson13} and directly compute the expectation of $\phi_{k}^{(g)}$ in a vectorized form for each group. 
\end{enumerate}
In cases where the model is fitted with the Beta-Bernoulli prior \eqref{eqn:beta-bernoulli}  instead of the logit construction \eqref{eqn:6}, Step 5 and 6 of the VB algorithm are replaced by a step to update the variational parameters $\{m^{(g)},n^{(g)}\}$ used to approximate $\pi^{(g)}$. Explicit expressions for the update rules can be found in the Appendix.

The VB algorithm is terminated when either when changes in the objective are less than a pre-specified threshold $\epsilon = 0.01$ or when it has reached the maximum number of iteration.  Another stopping rule is to monitor changes in the entropy of the selection parameter $\gamma_{k}^{(g)}$.  For posterior inference, variable selection can be performed by first estimating the marginal posterior probabilities (MPPs) of inclusion $p(\gamma^{(g)}_k| \cdot)$ via the values of the variational parameter $\nu_{k}^{(g)}$, computed as in \eqref{eqn:19}, at convergence. Non-zero effective connectivities at the group level can then be selected by thresholding these estimates at a pre-specified threshold.  Given the selected non-zero connectivities, estimates of their magnitude and directionality can be achieved via inference on the parameters $\omega_k^{(g)}$, for groups $g=1,\ldots, G$, estimated from the VB algorithm as the values of $\mu_{k}^{(g)}$ at convergence.  In the variable selection setting, \cite{Scott2010} have showed that spike-and-slab prior constructions of the type we use here provide correction for multiple testing.

\subsection{User-friendly MATLAB software {\it BVAR-connect}}
\begin{figure}
\centering
\includegraphics[width = 1.4in,height=2.0in]{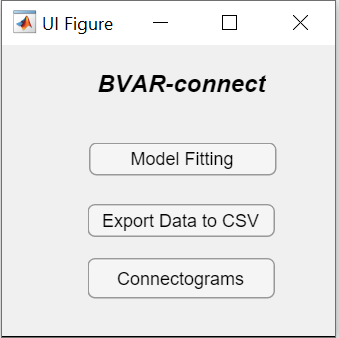}
\includegraphics[width = 2.4in,height=2.0in]{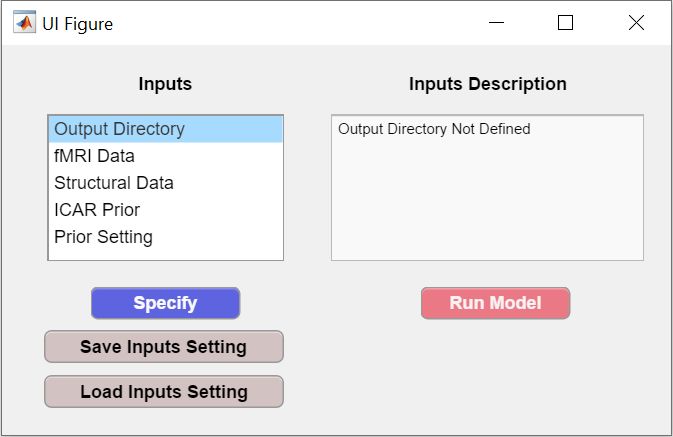}
\includegraphics[width = 2.2in,height=2.0in]{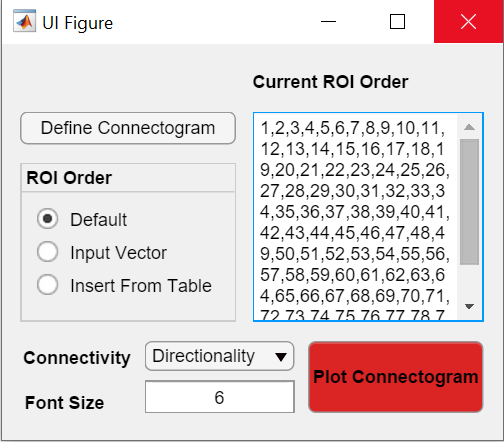}
\caption{{\it BVAR-connect} MATLAB GUI: Main GUI on the left, with links to interfaces to fit the model, export the output to a CSV file and visualize the output via connectograms. Model fitting interface in the center plot. Once all input arguments have been loaded/specified, the user can click on the Run button to fit the model. Visualization interface on the right. The order of the ROI's displayed in the connectograms can be defined by the user by clicking on the 'Insert From Table' option.}
\label{fig:interface}
\end{figure}

We give a brief description  of {\it BVAR-connect}, a MATLAB GUI that implements the Bayesian VAR model  with variational inference described in this paper. The GUI comprises two main interfaces, one for model fitting and one for visualization of the results, see Figure \ref{fig:interface}. It also allows users to export the output to a CSV file.

{\it Model fitting} interface: First, the user needs to load/specify a number of input parameters by selecting each of the objects in a listbox and clicking the Specify button:
\begin{itemize}
\item \textbf{Output Directory}: The user specifies the directory where the output will be stored. Once the model is run successfully, a \texttt{out.mat} file is generated in the output directory.
\item \textbf{fMRI Data}: The user loads a .mat file containing a $T\times R\times N$ array, named \texttt{X}, storing the fMRI times series data with $T$ time points, at $R$ regions, for $N$ subjects; a $1\times R$ cell object \texttt{ROI\_names}, storing the ROI's names, listed in the same order as stored in \texttt{X}; an integer \texttt{L} defining the number of lags of the BVAR model; an integer \texttt{G} specifying the number of groups, and \texttt{eta}, a $1\times N$ vector storing integer values ranging from $1$ to $G$, indicating which group each subject belongs to. 

\item \textbf{Structural Data}: The user loads a .mat file containing a $1\times G$ cell array object, \texttt{DTI\_vec}, where each cell contains a $(R^2\times L) \times 1$ vector storing the structural data. If a .mat file is not loaded, the model will run without any structural prior information.
\item \textbf{ICAR Prior}: The user loads a .mat file containing variable \texttt{S}, a $LR^2\times LR^2$ symmetric binary matrix used in the ICAR prior.
\item \textbf{Prior Setting}: The user is prompted to input the model hyperparameters from a window, which has pre-defined default settings for all parameters. 
\end{itemize}
Once all input arguments have been specified, the user can click on the Run button to fit the model, see right panel of Figure \ref{fig:interface}. If one chooses to run the model later, it is possible to press the \texttt{Save Inputs Setting} button to store the model specification, which can later be loaded using the \texttt{Load Inputs Setting} option.


{\it Visualization} interface: After running the model successfully, selected connectivities can be visualized as connectograms via the interface \texttt{Connectograms}. This interface,  shown in the right panel of Figure \ref{fig:interface},  can be used to draw simple connectograms that allow the user an initial exploration of the results, without the need to export the output. The GUI also allows to export the selected effective connectivities into a CSV file using the \texttt{Export Data to CSV} interface, so that the output can be visualized using one's preferred method or software (e.g., the {\it circlize} package in R). The visualization interface comprises the following arguments to be specified:

\begin{itemize}
\item \textbf{Define Connectogram}: Given the $G$ subject groups, the user defines the group to visualize. Connectivites can be filtered based on the remaining groups, as those shared by all groups, unique to a group, or any other possible combination. 

\item \textbf{ROI Order}: The user defines the ordering of the ROI's displayed in the connectograms, which by default are sorted based on the ordering in the \texttt{X} array.  The user can change the order either by inputing a vector via the \texttt{Input Vector} option or by clicking on \texttt{Insert From Table} and defining a new order manually. 

\item \textbf{Connectivity}: The user selects one of two possible options for plotting connectograms. The default option is \texttt{Directionality}, where edges are colored by the direction of the connectivity.  For example, if there exists an effective connectivity from region A to region B, the edge connecting the two regions will take the color representing region A. The other option is \texttt{Sign}, which defines the edge color based on the sign of the estimated connectivities, red for positive values and blue for negative.
\end{itemize}

\section{Simulation Studies}
\label{sec:simulation}

We first use simulated data to assess the accuracy of the variational approach with respect to the sampling-based MCMC method of \cite{Chiang17}.  The purpose of this comparison is to show that the VB algorithm reduces the computational cost without compromising the accuracy of the estimation.  We note that \cite{Chiang17}  already showed higher accuracy in the detection of effective connectivity at the group-level than competing approaches, such as those that use Granger causality and the VAR models of \cite{gorrostieta2012investigating,gorrostieta2013hierarchical}. We also test scalability of the VB algorithm on a larger data set, with unequal sample sizes, and discuss the sensitivity of its accuracy to the model hyperparameters.
	
\subsection{Performance comparison to MCMC}

We follow the data generating procedure of \cite{Chiang17} with minor changes in the parameter setting. In particular, we set the number of regions to $R=10$, the total sample size to $n = 20$ and the number of groups to $G=2$, with the first 10 subjects belonging to group 1 and the other 10 to group 2. We generated the data for each subject from the model described in Section~\ref{sec:methods}, using a VAR process of order $L = 1$ and with $T = 400$. We set the structural connectivity matrices to
	
\begin{equation}
\boldsymbol{N}^{(1)}=vec\left(\left[\begin{array}{cccccccccc}
0.1 & 0.3 & 0.1 & 0.6 & 0.2 & 0.1 & 0.7 & 0.3 & 0.1 & 0.4\\
0.3 & 0.5 & 0.1 & 0.1 & 0.2 & 0.1 & 0.6 & 0.1 & 0.5 & 0.15\\
0.1 & 0.1 & 0.35 & 0.65 & 0.2 & 0.7 & 0.1 & 0.6 & 0.2 & 0.4\\
0.6 & 0.1 & 0.65 & 0.3 & 0.25 & 0.1 & 0.2 & 0.15 & 0.4 & 0.3\\
0.2 & 0.2 & 0.2 & 0.25 & 0.1 & 0.85 & 0.3 & 0.25 & 0.1 & 0.15\\
0.1 & 0.1 & 0.7 & 0.1 & 0.85 & 0.5 & 0.3 & 0.1 & 0.25 & 0.2\\
0.7 & 0.6 & 0.1 & 0.2 & 0.3 & 0.3 & 0.1 & 0.05 & 0.12 & 0.3\\
0.3 & 0.1 & 0.7 & 0.15 & 0.25 & 0.1 & 0.05 & 0.25 & 0.3 & 0.15\\
0.1 & 0.5 & 0.2 & 0.4 & 0.1 & 0.25 & 0.12 & 0.3 & 0.1 & 0.3\\
0.4 & 0.15 & 0.4 & 0.3 & 0.15 & 0.2 & 0.3 & 0.15 & 0.3 & 0.2
\end{array}\right]\right)
\nonumber
\end{equation}
for group 1 and to
\begin{equation}
\boldsymbol{N}^{(2)}=vec\left(\left[\begin{array}{cccccccccc}
0.2 & 0.5 & 0.3 & 0.1 & 0.4 & 0.5 & 0.1 & 0.2 & 0.3 & 0.1\\
0.5 & 0.3 & 0.1 & 0.3 & 0.1 & 0.1 & 0.1 & 0.15 & 0.05 & 0.3\\
0.3 & 0.1 & 0.55 & 0.15 & 0.5 & 0.5 & 0.1 & 0.5 & 0.1 & 0.4\\
0.1 & 0.3 & 0.15 & 0.1 & 0.35 & 0.3 & 0.6 & 0.1 & 0.4 & 0.5\\
0.4 & 0.1 & 0.5 & 0.35 & 0.36 & 0.1 & 0.2 & 0.1 & 0.25 & 0.05\\
0.5 & 0.1 & 0.5 & 0.3 & 0.1 & 0.1 & 0.1 & 0.2 & 0.5 & 0.15\\
0.1 & 0.1 & 0.1 & 0.6 & 0.2 & 0.1 & 0.7 & 0.25 & 0.4 & 0.2\\
0.2 & 0.15 & 0.5 & 0.1 & 0.1 & 0.2 & 0.25 & 0.25 & 0.3 & 0.15\\
0.3 & 0.05 & 0.1 & 0.4 & 0.25 & 0.5 & 0.4 & 0.3 & 0.1 & 0.25\\
0.1 & 0.3 & 0.4 & 0.5 & 0.05 & 0.15 & 0.2 & 0.15 & 0.25 & 0.3
\end{array}\right]\right)
\nonumber
\end{equation}
for group 2, and generated random matrices $A^{(s)}=Q'(s)\Lambda Q^{(s)}$, with a diagonal matrix $\Lambda$ set to $\Lambda=Diag(-0.4, -0.25,-0.1, 0.05$  $,0.2,-0.3, 0.1, 0.1,-0.3,-0.15)$ and $Q^{(s)}$ a randomly generated orthogonal matrix from the QR decomposition of a matrix of standard normal random deviates. The subject-level connectivities were then obtained as
\begin{equation}
\beta^{(s)}_{g}\sim\left(\Omega^{(1)}+vec(A^{(s)})\right)1_{[1\geq s\geq 10]}+
\beta^{(s)}_{g}\sim\left(\Omega^{(2)}+vec(A^{(s)})\right)1_{[11\geq s\geq 20]}.\nonumber
\end{equation}
The purpose of using a generating mechanism of the subject-level deviations from the group connectivities, which is different from the assumed prior model, is to test whether the model can correctly estimate group-level connectivities in a robust manner.

\subsection{Parameter settings}

We fitted the proposed Bayesian VAR model with lag $L=1$. Noninformative priors were used, largely following \cite{Chiang17}, that is, we set $h_1=2$ and $h_2=1$ for the variance parameter $\zeta_j$, and $a_0^{(g)}=a_1^{(g)}=2, b_0^{(g)}=b_1^{(g)}=1$ for $g=1,2$ for the variance parameters on the subject-level connectivities. In the ICAR prior we fixed $q$ to a large value (i.e. $>50$) to obtain a vague specification and set $S$ to encourage smoothness among connectivities at a same lag that initiate from the same node. As for our probit prior, we set $\alpha^{(g)}_0=-2.944$,  which gives a prior inclusion probability of 0.05 when external information is not present, and fixed the prior mean and variance of $\alpha_1^{(g)}$ to $w^{(g)}=0$ and $\tau^{2(g)}=100$, for $g=1,2$. 

MCMC chains were run with 40,000 iterations using 10,000 sweeps as burn-in. Convergence was assessed by using the Raftery–Lewis' and Geweke's diagnostic criteria implemented in the R package ``coda". The VB algorithm requires initial values for the parameters of the variational distributions in equation \eqref{eqn:14}, except for the $\underline{\beta}_{g}^{(s)}$'s, which are updated at the first step of the algorithm. We initialized means and variances of the variational distribution modeling $\tilde{\omega}
 _{k}^{(g)}$ by sampling the means from $\mu_{k}^{(g)}\sim Unif(-0.5,0.5)$ and setting $ s_{k}^{2(g)}=10$. Next, we initialized the distributions of the variance parameters of effective connectivity as $q(\xi_{1}^{(g)})\sim IG(2,20)$ and $q(\xi_{0}^{(g)})\sim IG(2,10)$. The error terms of the model were modeled as $q(\zeta_{j})\sim IG(2,5)$, for $j=1,\ldots,R$. Lastly, the model selection variable $\gamma_{k}^{(g)}$'s were initialized as $q(\gamma_{k}^{(g)})\sim Bernoulli(\nu_{k}^{(g)}=0.1)$.  Finally, the \Polya-Gamma augmentation scheme requires initial values for the parameters of the variational distribution of $\alpha^{(g)}_1$. We have found that a group-specific setting $(\mu_{\alpha_{1}^{(g)}}, \sigma_{\alpha_{1}^{(g)}}^2) =(C\times S_{g}/\bar{N}^{(g)}, 10)$, with $\bar{N}^{(g)}$ the mean of the structural connectivity matrix $N^{(g)}$ for group $g$ across all regions and $C$ a constant chosen in the range $[50,100]$, works well, particularly in situations with sample groups with unbalanced sample sizes (see simulation study in Section 3.5 for more details).
	
Simulations were run on an Intel Xeon E5-2630 station (2.30GHz) with 132 GB RAM.  Non-zero effective connectivities at group level were selected by thresholding the estimated $\nu_{k}^{(g)}$, computed as in \eqref{eqn:19}, at the value .5, resulting in the median model. 
	
\subsection{Results}	
Table~\ref{table:comparison} reports false positive rate (FPR), false negative rate (FNR), accuracy and the $F_{1}-$score, averaged over 30 replicated simulated datasets. Furthermore, mean squared errors (MSE) between the estimated and the true connectivities, calculated as
\begin{equation*}
MSE=\frac{(vec(\underline{B}_{g})- vec(\underline{\hat B}_{g}))^T (vec(\underline{B}_{g})- vec(\underline{\hat B}_{g}))}{LR^2},	
\end{equation*}
and averaged over 30 replicates, were $[0.0002,0.0003]$ for the MCMC approach and $[0.0002, 0.0004]$ for the VB approach, for groups 1 and 2, respectively. Results show performances of the VB approach quite comparable to the MCMC sampling-based method. Overall, the MCMC approach results in a more conservative selection of non-zero connectivities, relative to the VB approach, by small margins, as shown by the lower FPRs and higher FNRs. The higher Accuracy and $F_{1}$-scores of the VB model can be attributed to its low FNRs.  
	
As expected, the VB approach outperformed the MCMC method in terms of computational cost. In particular, on a single replicated dataset, $40,000$ iterations of the MCMC algorithm took $40$ hours, whereas the VB approach converged within a minute after running less than $100$ iterations. These results confirm that the VB algorithm works well for our modeling setting, as it is able to reach estimation accuracy comparable to sampling-based techniques, while reducing the computational cost. Furthermore, performances of our model were consistent for higher lags. In particular, repeating the simulation study with $L=2$ resulted in FPR=0.0009, FNR=0.2823, accuracy=0.8717 and $F_1=0.9242$, for group 1, and  FPR=0.0041, FNR=0.2812,  accuracy=0.8685 and $F_1=0.8902$, for group 2.

\begin{table}
\centering 
\begin{tabular}{c c c c} 
\hline\hline 
&  & MCMC & VB\\ [0.5ex] 
\hline 
Group 1 & FPR & 0.0113 & 0.0196   \\
& FNR & 0.2207 & 0.1527 \\ 
& Accuracy & 0.9024 & 0.9250 \\ 
& $F_{1}$-score & 0.866 & 0.9032 \\
Group 2 & FPR & 0.0047 & 0.0239 \\
& FNR & 0.2205 & 0.1274 \\ 
& Accuracy & 0.8714 & 0.9343  \\ 
& $F_{1}$-score & 0.9087 & 0.9141  \\ [1ex] 
\hline 
\end{tabular}
\caption{Simulated data with $R=10$ regions: Performance comparison of VB and MCMC algorithms. Results are averaged over 30 simulated data. } 
\label{table:comparison} 
\end{table}
		
\subsection{Sensitivity analysis}
We also investigated performances of the VB algorithm in a larger simulated setting, with $R=30$ regions, $G=2$ and $n=80$, where we assigned the first $20$ subjects to group $1$ and the other $60$ to group $2$.
This scenario is more similar  to the case study, in particular in its characteristic unequal sample size for the two groups. We generated the data using a similar procedure to the one above. We set $L=1$ and $T=150$. We generated structural connectivity matrices as follows:
\begin{enumerate}
\item Define a $30\times30$ zero matrix $N^{(i)}$ for $i=1,\ldots G$.
\item Fill the upper triangular part of $N^{(i)}$ with random variables sampled from a uniform distribution with support $(0.3,0.7)$.
\item Out of the $465$ non-zero entries, randomly sample $400$ indexs and replace original values to $0.1$.
\item We add the transpose of $N^{(i)}$ excluding the diagonal elements to itself.
\item We reset the diagonal elements of $N^{(i)}$ to  $min(diag(N^{(i)})+0.5,1)$.
\end{enumerate}
We defined a $30\times30$ diagonal matrix $\Lambda$, with diagonal entries sampled from a uniform distribution with support $(-0.4,0.3)$, and set the parameters of the logistic prior in \eqref{eqn:6} to $\alpha_{0}=-2.5$ and $\alpha_{0} =5$, for both groups.  This process resulted in a dataset where approximately 10 percent of the group effective connectivities had non-zero values. 
	
Table~\ref{table:comparison2} provides summaries over 30 replicated datasets of the performance measures for different values of the prior parameter $\alpha_{0}$, which controls the prior probability of observing non-zero effective connectivities when structural connectivity is not present. The chosen values correspond to prior probabilities $p(\gamma_{k}^{(g)}=1\mid N_{k}^{(g)}=0)$ of $[0.02,0.05,0.1]$.   In the last column of the table, to better understand the effect of incorporating structural data into the model, we report results obtained by fitting the model with the Beta-Bernoulli prior \eqref{eqn:beta-bernoulli} with $e^{(g)}=0.1,f^{(g)}=1.9$, which gives an equal prior inclusion probability of $0.05$.  Results show that modest changes in $\alpha_{0}$ do not affect the accuracy of the estimation. They also show that including the structural data into the prior contributes to improving FPR and FNR, consequently leading to higher $F_{1}-$scores. One thing to notice is that we observe a much lower FNR values for group 2. This can be attributed to the larger sample size for this group, which leads to higher selections of true non-zero effective connectivities. 
In terms of computation time, the VB algorithm converged after 50 iterations, which took about 15 minutes, for an individual dataset. Similarly to $\alpha_{0}$, we found that small changes in the other model hyperparameters did not affect the accuracy of our VB estimation.
	
\begin{table}
\centering 
\begin{tabular}{c c c c c|c} 
\hline\hline 
& &$\alpha_{0}=-4$ &$\alpha_{0}=-2.9$& $\alpha_{0}=-2.2$ & Beta(0.1,1.9)\\ [0.5ex]
\hline
Group 1 & FPR & 0.0001 & 0.0001 &0.0001 & 0.0009\\
& FNR & 0.3581 & 0.3649 &0.3752 & 0.3731\\ 
& Accuracy & 0.9921 & 0.9920& 0.9918 & 0.9910\\ 
& $F_{1}$-score & 0.7753 & 0.7707 & 0.7629 & 0.7483\\
Group 2 & FPR & 0.0011 & 0.0008& 0.0006 & 0.0024\\
& FNR & 0.1511 & 0.1825 &0.1763 & 0.1967\\ 
& Accuracy & 0.9953 & 0.9954& 0.9953 & 0.9931\\ 
& $F_{1}$-score & 0.8923 & 0.8937&0.8890 & 0.8386\\ [1ex] 
\hline 
\end{tabular}
\caption{Simulated data with $R=30$ regions: Results are averaged over 30 simulated datasets, for varying values of the parameter $\alpha_{0}$} 
\label{table:comparison2}
\end{table}

\subsection{Example with large simulated network}

In order to mimic the size of our case study, and to show scalability of our method, we also investigated performances of the VB algorithm with a network of $R=90$ regions. We set $G=2$ and $n=100$, with subjects equally split between the two groups. We set $L=1$ and $T=150$. We generated structural connectivity matrices similarly to what described for the smaller network cases above,  imposing the amount of connectivity to be about 15\% of the total number of edges. Results averaged over 30 replicated datasets resulted in FPR= 0.0, FNR=0.5279, accuracy=0.9169 and $F_1=0.6412$, for group 1, and  FPR=0.0, FNR=0.5274,  accuracy=0.9176 and $F_1=0.6418$, for group 2. Lower FNRs can be obtained with larger sample sizes. For example, $n=200$ led to FPR= 0.0, FNR=0.3677, accuracy=0.9421 and $F_1=0.7747$, for group 1, and  FPR= 0.0, FNR=0.3670,  accuracy=0.9426 and $F_1=0.7752$, for group 2.

\subsection{VB Initialization in case of unequal sample sizes}
Our proposed variational algorithm is mostly deterministic. Hence, different initializations of $\Theta$ can affect the accuracy of the estimation. In particular, in simulations we noticed that the initial choice of the mean parameter $\mu_{\alpha_{1}^{(g)}}$ of $\alpha_{1}^{(g)}$ largely influenced the estimation of the model selection variable $\gamma_{k}^{(g)}$. The underlying reason for this can be understood by looking at the update rule of the log odds $\rho_{k}^{(g)}$ of $\gamma_{k}^{(g)}$ shown in Appendix \ref{app:BVAR_VI}. As demonstrated in the first line of the update rule for $\rho_{k}^{(g)}$, the posterior probability of observing non-zero effective connectivities for group $g$ is inversely proportional to $S_{g}$, the number of subjects belonging to that group. So, by default, the group with more subjects will have a sparser effective connectivity. If the variational distribution of $\alpha_{1}^{(g)}$ is not initialized properly to counter this negative weight, $\rho_{k}^{(g)}$ will be fixed at a large negative value, leading to no variable selection and suboptimal convergence of the VB algorithm. We have found that a group-specific initialization of the type $\mu_{\alpha_{1}^{(g)}}=C\times S_{g}/\bar{N}^{(g)}$, with $\bar{N}^{(g)}$ the mean of the structural connectivity matrix $N^{(g)}$ for group $g$ across all regions and $C$ a constant chosen in the range $[50,100]$, does address the imbalance issue. A similar issue occurs in the model without external structural data, for the initialization of the pairs $\{m^{(g)},n^{(g)}\}$ modeling $\pi^{(g)}$. For this case, we found the initialization setting $m^{(g)}=3,n^{(g)}=0.005$ to resolve the imbalance issue. With these initial values, the VB algorithm starts from a full model, gradually performing variable selection in $\gamma_{k}^{(g)}$ and converging to a higher ELBO objective. Even though ad-hoc, we have found these suggested rules to work well in a variety of simulated scenarios and in the analysis of data from different studies. As far as the remaining variational parameters were concerned, we found that the model accuracy was not affected by modest changes of the initial values.

\section{Case Study on Traumatic Brain Injury}
\label{sec:fMRI}
We illustrate our method on resting-state functional MRI and structural DTI data collected on children with a history of traumatic injury.
	
\subsection{Experimental study}
We obtained data from a prospective, longitudinal study in children with a history of traumatic injury following a vehicle collision, recruited from the Emergency Department or Level 1 Pediatric Trauma Center at Children's Memorial Hermann Hospital/University of Texas Health Science Center at Houston (UTHealth) between September 2011 and August 2015. 
	
Participants met the following inclusion criteria: 1) injured in a vehicle accident between 8 and 15 years of age; 2) proficiency in English or Spanish; 3) residing within a 125 mile catchment radius; 4) no prior history of major neuropsychiatric disorder (intellectual deficiency or low-functioning autism spectrum disorder) that would complicate assessment of the impact of injury on brain outcomes; 5) no metabolic, endocrine, or systemic health problems (e.g., hypertension); 6) no prior medically-attended TBI; and 7) no habitual use of steroids, tobacco, or alcohol. The latter four criteria were assessed during screening using a brief parent interview. Participants were further classified into subgroups reflecting either injury to the head (i.e., traumatic brain injury, TBI) or injury to the body with no direct impact to the head (i.e., extra cranial injury, EI).  Of 220 injured youth who met study inclusion criteria, 131 were consented and enrolled, and 113 were scanned at baseline (TBI: n=81; EI: n=34). 
A quality control evaluation of all scans resulted in 70 TBI and 27 EI samples being selected for analysis. Subjects were excluded for excessive motion or scanner error (e.g., operator error, crack in scanner head coil). TBI severity was determined by the lowest post-resuscitation Glasgow Coma Scale (GCS) \cite{Teasdale1974} score and by acute neuroimaging findings. 
Written informed consent was obtained from each child's guardian and written assent was obtained from all children in accordance with Institutional Review Board guidelines. 

For this investigation, children with history of EI were included as a comparison group, rather than a healthy non-injured sample, to account for characteristics such as increased risk-taking behaviors that might be elevated in children who are most likely to sustain injury. The benefit of comparison between children with TBI and EI group is that differences in group effective connectivity likely correspond to specific injury status.

\subsection{Data preprocessing}
MRI data were acquired on a Philips 3 Tesla (T) Intera system with a 32-channel head coil at the University of Texas McGovern Medical School. The T1-weighted (T1W) sequence was acquired in the sagittal plane with parameters: TR/TE = 8.07 / 3.68 ms; flip angle = 6$^o$; acquisition matrix = 256 x 256; FOV = 256 mm; slice thickness = 1 mm; resulting in a voxel size = 1 x 1 x 1 mm$^3$. The DTI sequence was acquired using single-shot spin-echo echo planar imaging (EPI) with parameters: TR/TE = 8700 / 67 ms; flip angle = 90$^o$; FOV = 240 mm; matrix size = 96 x 96; resulting in a voxel size = 2.5 x 2.5 x 2.5 mm$^3$. A single non-diffusion weighted volume was acquired (b = 0 s/mm$^2$), along with 32 non-collinear diffusion-weighted volumes (b = 1000 s/mm$^2$) distributed uniformly. Resting-state fMRI data were acquired in the axial plane using gradient-echo EPI with parameters: TR/TE = 2000 / 30 ms; flip angle = 90$^o$; FOV = 240 mm; slice thickness = 2.5 (skip 1) mm; acquisition matrix = 192 x 192; resulting in a voxel size =  1.25 x 1.25 x 3.50 mm$^3$. Subjects were instructed to rest quietly with eyes closed for the duration of the sequence; 180 volumes were acquired with acquisition time = 6:06. At approximately the mid-point of data collection, the scanner was upgraded to a Philips 3T Ingenia system. Fidelity analysis was performed to match fMRI, diffusion-weighted and T1-weighted scanning protocols. Although some variability might remain between pre- and post-scanner upgrade, distribution of scanner upgrade was similar between groups as determined by ANOVA (p=.37).
	
Resting state fMRI data were preprocessed using SPM12 (Wellcome Trust Center for Neuroimaging, http://www.fil.ion.ucl.ac.uk/spm/). Preprocessing steps included motion correction through realignment; slice timing correction; segmentation of gray matter, white matter, and CSF; registration to the subject's T1-weighted MPRAGE image; registration to a standard space using the ICBM space template for European Brains; and smoothing using a full-width half maximum (FWHM) Gaussian kernel of 8mm. Resulting images were entered into the Artifact Detection Tools (ART) toolbox \cite{whitfield2011artifact} to identify and exclude volumes with extreme levels of motion (greater than 2mm in any of the six dimensions - x, y, z, pitch, roll, yaw). Subjects with motion outliers in more than 15\% of the volumes were removed from analyses. Prior to the analysis, a 3D parcellation was performed on the preprocessed data using the MarsBaR toolbox in SPM 12. For parcellation, the automatic anatomical labelling (AAL) brain atlas was used, which resulted in 90 ROI's excluding regions associated with the cerebellum.
	
Preprocessing of DTI data was performed with FSL and the FSL Diffusion Toolbox (FDT) \cite{behrens2003characterization}. A total of 22 subjects did not have usable DTI data and were excluded from DTI analysis (TBI: 7 female, 12 male; EI: 2 female, 1 male). The following steps were applied to each study participant individually. First, eddy-current correction was performed using the latest GPU version of eddy to correct for image distortions and head motion \cite{andersson2003correct,andersson2015non,andersson2016integrated,andersson2016incorporating,graham2016realistic}. Fractional anisotropy (FA) was calculated using FSL's DTIFIT to measure white matter tract integrity. Tract based spatial statistics (TBSS), a fully automated voxel based approach to DTI analysis, was used to align FA maps from each subject with useable DTI data to the FA map of the most representative subject of their respective group.  Each subject's mean FA skeleton was thresholded at 0.20 to restrict estimation to white matter tracts and 
 exclude partial volume effect from neighboring gray matter. For each of the AAL regions, white matter tracts of interest were masked bilaterally and symmetrically. Resulting masks were overlaid over individual white matter tract skeletons to provide FA values for the tracts of interest for each subject. 
	
\subsection{Results}
We fitted the proposed Bayesian VAR model with lag $L=1$. Typically, with fMRI data, VAR models of order one or two are recommended, given the low temporal resolution of the data. Diagnostics are often based on auto-correlation of the residuals. Also, criterion functions such as the Schwarz's Bayesian information criterion (BIC) \cite{schwarz1978} can be used. Here we use $L=1$, as this lag is shown to be appropriate for detection of both positive and negative connections within human fMRI data. For example, \cite{Goelman2014} show how positive and negative connections dissociate with increasing lag-time, that is, positive connections become less significant with increasing lag whereas negative connections increase, in both the human and rats.  We refer readers to \cite{Chiang17} for our application of the model with $L=2$.

For VB, initialization of  the variational distributions and hyperparameters setting were done as described in Section 3.2. Results we report here were obtained by performing the analyses on an Intel Xeon E5-2630 station (2.30GHz) with 132 GB RAM. The VB algorithm converged after 25 iterations, with an elapsed time of approximately 8 hours.  Non-zero effective connectivities were selected by thresholding the estimated $\nu_{k}^{(g)}$, computed as in \eqref{eqn:19}, at the value .5. When running the analyses we noticed that increasing the number of iterations would result in more and more separation between selected and non-selected edges, with estimated $\hat\nu_{k}^{(g)}$ getting closer to the one and zero values. This behavior was consistent across all simulations and real data analyses.
	
\begin{figure}
\centering
\includegraphics[width =4.4in,height=4.4in]{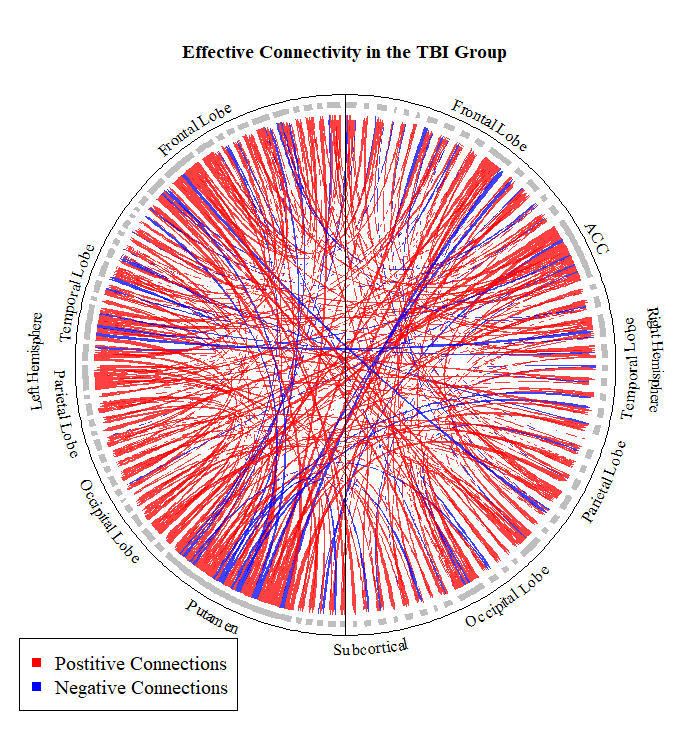}
\includegraphics[width =4.4in,height=4.4in]{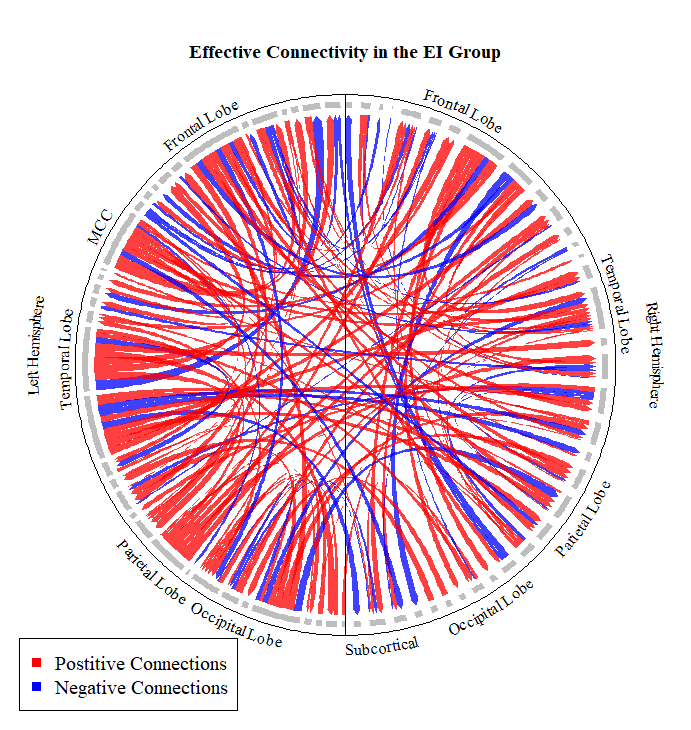}
\caption{TBI data: Connectograms of the estimated effective connectivities unique to the TBI and EI groups. Red edges denote positive values and blue edges denote negative values. }
\label{fig:networks_TBIEI}
\end{figure}

\begin{figure}
\centering
\includegraphics[width =4.4in,height=4.4in]{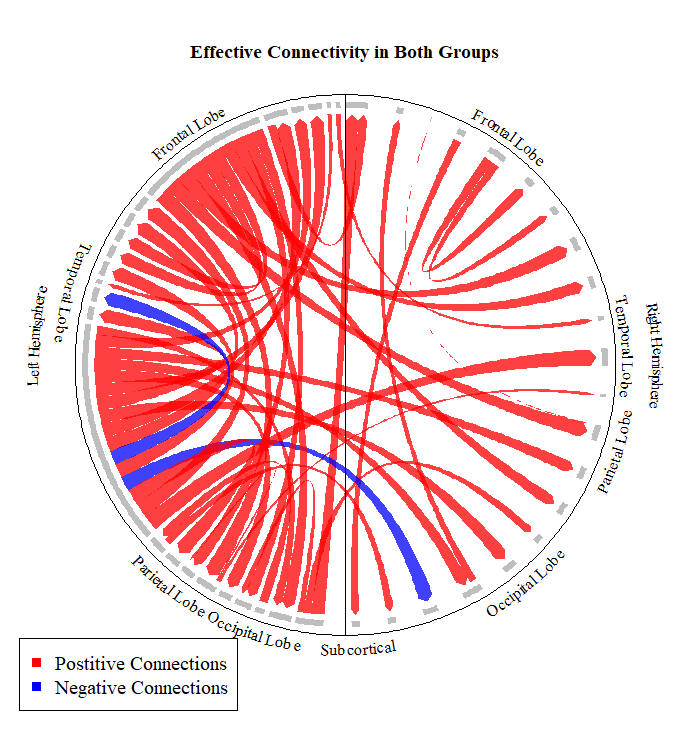}
\caption{TBI data: Connectogram of the estimated effective connectivities shared by the two groups. Red edges denote positive values and blue edges denote negative values. }
\label{fig:networks_shared}
\end{figure}

Estimated brain connectivities are summarized in the connectograms shown in Figures \ref{fig:networks_TBIEI} and \ref{fig:networks_shared}. Unique edges to the TBI and EI groups are shown in Figure \ref{fig:networks_TBIEI} and shared edges between the two groups in Figure \ref{fig:networks_shared}.  These connectograms were obtained using the {\it circlize} package in R \cite{Gu2014}. Positive connections between regions are shown in red, and negative connections are shown in blue.  The R script and spreadsheets used to create these connectograms are publicly available at  \url{https://osf.io/c4qgw/}. Numbers of unique and shared connections within and between brain lobes in the two groups are reported in Tables \ref{tab:unique} and \ref{tab:shared}, respectively, together with estimated mean strengths and standard deviations.  For each group, the results were formatted in terms of the presence/absence of a connection as well as the strength of the connections that were present. Only connections that were unique to each group were included in the Tables and these connections were categorized based on the region from which the edge begins (i.e., the time 1 region). Categories were determined based on lobe (frontal, limbic, occipital, parietal, subcortical, temporal). Results in the tables show that the TBI group has more unique connections than the EI group; however, the unique connections of the EI group are stronger relative to the TBI group. 	

\begin{table}
\centering
\begin{tabular}{|c|c|c|c|c|}
\hline
\multicolumn{5}{|l|}{\textit{\textbf{Unique Connections}}}      \\ \hline
\textbf{Lobe} & \textbf{Hemisphere}& \textbf{Group} & \textbf{\begin{tabular}[c]{@{}c@{}}Number of Unique\\ Connections\end{tabular}} & \textbf{\begin{tabular}[c]{@{}c@{}}Strength of Unique\\ Connections (mean/std)\end{tabular}} \\ \hline
\multirow{2}{*}{\textbf{Frontal}}     & \multirow{2}{*}{Left, Right} & EI  & 17,16   & 0.33(0.10), 0.35(0.11)\\ 
\cline{3-5} &   & TBI   & 39,23  & 0.29(0.16), 0.22(0.11) \\ \hline
\multirow{2}{*}{\textbf{Limbic}}   & \multirow{2}{*}{Left, Right} & EI   & 16,0& 0.29(0.12), NA    \\ \cline{3-5} &   & TBI&1,30&0.02(NA), 0.31(0.13) \\ \hline
\multirow{2}{*}{\textbf{Occipital}}   & \multirow{2}{*}{Left, Right} & EI  & 12,6 & 0.38(0.12), 0.33(0.15)  \\ \cline{3-5}    &     & TBI  &17,28&0.28(0.09), 0.24(0.10)\\ \hline
\multirow{2}{*}{\textbf{Parietal}}    & \multirow{2}{*}{Left, Right} & EI    & 6,1  & 0.39(0.05), 0.16(NA)   \\ \cline{3-5}    &     & TBI    &17,18&0.18(0.07), 0.16(0.08) \\ \hline
\multirow{2}{*}{\textbf{Subcortical}} & \multirow{2}{*}{Left, Right} & EI   & 2,0   & 0.23(0.02), NA \\ \cline{3-5}   &     & TBI  &51,1&0.39(0.130), 0.11(NA) \\ \hline
\multirow{2}{*}{\textbf{Temporal}}    & \multirow{2}{*}{Left, Right} & EI & 20,4  & 0.45(0.10), 0.23(0.09) \\ \cline{3-5} &  & TBI & 16,14  & 0.36(0.11), 0.21(0.08) \\ \hline
\end{tabular}
\caption{TBI data: Numbers of unique connections within each lobe of the brain, for the TBI and EI groups, together with estimated mean strengths and standard deviations. } 
\label{tab:unique}
\end{table}

\begin{table}
\centering
\begin{tabular}{|c|c|c|c|c|}
\hline
\multicolumn{5}{|l|}{\textit{\textbf{Shared Connections}}}      \\ \hline
\textbf{Lobe} & \textbf{Hemisphere}& \textbf{Group} & \textbf{\begin{tabular}[c]{@{}c@{}}Number of Shared\\ Connections\end{tabular}} & \textbf{\begin{tabular}[c]{@{}c@{}}Strength of Shared\\ Connections (mean/std)\end{tabular}} \\ \hline
\multirow{2}{*}{\textbf{Frontal}}     & \multirow{2}{*}{Left, Right} & EI  & 13,3   & 0.48(0.13), 0.38(0.01)\\ 
\cline{3-5} &   & TBI   & 13,3  & 0.38(0.15), 0.30(0.08) \\ \hline
\multirow{2}{*}{\textbf{Limbic}}   & \multirow{2}{*}{Left, Right} & EI   & 0,0& NA, NA    \\ \cline{3-5} &   & TBI&0,0& NA, NA \\ \hline
\multirow{2}{*}{\textbf{Occipital}}   & \multirow{2}{*}{Left, Right} & EI  & 1,1 & 0.29(NA), 0.23(NA)  \\ \cline{3-5}    &     & TBI  &1,1&0.41(NA), 0.33(NA)\\ \hline
\multirow{2}{*}{\textbf{Parietal}}    & \multirow{2}{*}{Left, Right} & EI    & 1,0  & 0.46(NA), NA   \\ \cline{3-5}    &     & TBI    &1,0&0.11(NA), NA \\ \hline
\multirow{2}{*}{\textbf{Subcortical}} & \multirow{2}{*}{Left, Right} & EI   & 3,0   & 0.43(0.25), NA \\ \cline{3-5}   &     & TBI  &3,0&0.26(0.04), NA \\ \hline
\multirow{2}{*}{\textbf{Temporal}}    & \multirow{2}{*}{Left, Right} & EI & 16,0  & 0.58(0.10), NA \\ \cline{3-5} &  & TBI & 16,0  & 0.45(0.12), NA \\ \hline
\end{tabular}
\caption{TBI data: Numbers of shared connections between brain lobes, for the TBI and EI groups, together with estimated mean strengths and standard deviations. }
\label{tab:shared}
\end{table}

Previous research comparing children with TBI to children with EI has found that children with TBI have decreased white matter connectivity in core pathways of the brain (as measured by fractional anisotropy \cite{Ewing16,watson19,wilde12}. Consistent with structural findings, our results suggest that functional connectivity across regions may also be weaker in children with TBI, although they may have a larger number of connections. The increased number of connections in children with TBI could indicate that the brain may be using compensatory functional connections in response to damage to other functional and/or structural connections. To understand this further, it is important to consider the location of the connections (i.e., hemisphere and lobe). Overall, left hemisphere connections were stronger than right hemisphere connections. The EI group had stronger connections than the TBI group in the frontal lobe, occipital lobe, parietal lobe, and temporal lobe, but the groups did not differ in the strength of the connections in limbic regions or subcortical regions.

\section{Conclusions}\label{sec:discussion}
In this paper, we have developed {\it BVAR-connect}, a variational approach to a multi-subject Bayesian vector autoregressive model proposed by \cite{Chiang17} for inference on effective brain connectivity based on resting-state functional MRI data. Our framework uses logit priors for the integration of multi-modal data, in particular structural DTI data, and \Polya-Gamma augmentation schemes for tractable inference. The VB approach we have developed allows scalability of the methods and results in the ability to estimate brain connectivity networks over whole-brain parcellations of the data.  We have illustrated the methods on resting-state fMRI and DTI data on children with a history of traumatic injury.  Through simulations, we have shown that the VB algorithm attains comparable performance to sampling-based MCMC approaches at a significantly faster computation time.  We have also addressed the case of subject groups with imbalanced sample sizes.  

The model formulation we have adopted assumes that the hemodynamic response function has been modeled during the pre-processing of the data, for example via SPM. For the fMRI data presented in the current study, we assumed a canonical hemodynamic response function across all voxels of the brain. In SPM, researchers have the choice of including time and dispersion derivatives to account for some of the variability in the hemodynamic response function \cite{henson2001}. Future work on {\it BVAR-connect} could include extensions that incorporate the HRF into the model. Furthermore, extending VAR models and other modeling frameworks to the estimation of dynamic connectivities is an area of great interest in fMRI research \cite{Chang2010,Calhoun2014,Ryali2016,Chiang2018,Warnick2018}.

For the TBI investigation, we have considered children with history of EI as a comparison group, rather than a healthy non-injured sample, and have focused on differences in group effective connectivity, as those likely correspond to specific injury status.  However, both TBI and EI groups have a history of traumatic injury and thus it is unclear if shared functional connections reflect healthy brain functioning or a connectivity profile of children with history of traumatic injury more broadly. Indeed, previous research indicates reduced white matter connectivity between the hippocampus and the prefrontal cortex \cite{ewing2019post} in children with history of traumatic injury compared to healthy controls that was absent for the within injury comparison of EI vs. TBI.  Future investigations including a healthy comparison group could help in identifying neurodevelopmental effects of traumatic injury.

\section*{Information Sharing Statement}
The user-friendly MATLAB software {\it BVAR-connect} is available for download at 
\url{https://github.com/marinavannucci} and at 
\url{https://github.com/rimehi}. Detailed instructions on how to use the toolbox can be found in the Instructions text file. The visualization interface provided in the GUI is built upon the Matlab {\it circularGraph} package of Paul Kassebaum (Copyright 2016, The MathWorks, Inc. All rights reserved). No additional MATLAB toolboxes are required to run the software.

\section*{Acknowledgments}
Funding/support for this research was provided by (1) National Science Foundation NSF/SES 1659925 (MV); (2) Hamill Innovation Award, Rice University (MV and DDM); (3) National Institutes of Health R01 NS046308 (LEC). 

\section*{Compliance with Ethical Standards}
The content is solely the responsibility of the authors and does not necessarily represent the official views of the granting institutes. The authors report no financial or other conflict of interest.

\bibliographystyle{elsarticle-num}
\bibliography{myref}
\clearpage

\begin{appendix}
\section*{Appendix}
\label{app:BVAR_VI}
Our proposed variational algorithm consists of repeating the following steps to update the set of variational parameters $\Theta$ until convergence of the ELBO is met:
\begin{enumerate}
\item \textbf{Update $\{U_{g}^{(s)},\Sigma_{g}^{(s)}\}$ of $\underline{\beta}_{g}^{(s)}$, for $s=1,\ldots, n$}: We exploit conjugacy of the model and the approximating variational distribution, arriving to a closed-form expression for $\{U_{g}^{(s)},\Sigma_{g}^{(s)}\}$, with
\begin{eqnarray*}
U_{g}^{(s)} &=& \left(\mathbb{E}[\Xi^{-1}]\otimes(U^{T(s)}U^{(s)})+\mathbb{E}[\Sigma^{-(g)}]\right)^{-1}\left(\left(\mathbb{E}[\Xi^{-1}]\otimes U^{(s)}\right)\underline{x}^{(s)}+\mathbb{E}[\Sigma^{-(g)}]\mathbb{E}[\Omega^{(g)}]\right)\\
\Sigma_{g}^{(s)}&=&\left(\mathbb{E}[\Xi^{-1}]\otimes(U^{T(s)}U^{(s)})+\mathbb{E}[\Sigma^{-(g)}]\right)^{-1}.
\end{eqnarray*}
\item \textbf{Update $\{z_{1j},z_{2j}\}$ of $\zeta_{j}$, for $j=1,\ldots, R$}:  By conjugacy, closed form expressions can be obtained for $\{z_{1j},z_{2j}\}$. The update is performed in a vectorized manner, such that the parameters are updated jointly for all $j=1,\ldots, R$. For all $j$, $z_{1j}=\frac{N(T-L)}{2}+ h_{1}$. To update $z_{2j}$, we need to compute the following three terms for all subjects $s=1,\ldots, N$:
\begin{eqnarray*}
\underbrace{M_{1}^{(s)}}_{R-by-R} &=& \frac{1}{2}X^{(s)T}X^{(s)}\\
\underbrace{M_{2}^{(s)}}_{RL-by-R}&=&\frac{1}{2}vec^{-1}\left(diag\left((I\otimes(U^{(s)T}U^{(s)}))\times (U_{g}^{(s)T}U_{g}^{(s)}+\Sigma_{g}^{(s)})\right),RL,R\right) \\
\underbrace{M_{3}^{(s)}}_{RL-by-R}&=&vec^{-1}\left(\left(\underline{x}^{(s)T}(I\otimes U^{(s)})\right)\circ U_{g}^{(s)T},RL,R\right),
\end{eqnarray*}
where $I$ is a $R-by-R$ identity matrix, $vec^{-1}(V,M,N)$ an inverse vec operator that transforms a vector $V$ into a $M-by-N$ matrix, and $\circ$ the element-by-element (Hadamard) product. Then the vector $z_{2}$ can be updated as
\begin{equation*}
z_{2}= h_{2} + \sum_{s=1}^{n}\left[M_{1}(:,)^{(s)}+\frac{1}M_{2}(:,)^{(s)}-M_{3}(:,)^{(s)}\right],
\end{equation*}
where$M(:,)$ denotes the row sum of each matrix.
\item \textbf{Update $\{c_{1}^{(g)},d_{1}^{(g)}\},\{c_{0}^{(g)},d_{0}^{(g)}\}$ of $\xi_{1}^{(g)},\xi_{0}^{(g)}$ for $g=1,\ldots G$}:
By exploiting conjugacy, the pairs $\{c_{1}^{(g)},d_{1}^{(g)}\},\{c_{0}^{(g)},d_{0}^{(g)}\}$ are updated one group at a time as
\begin{eqnarray*}
c_{1}^{(g)} &=& \frac{S_{g}}{2}\sum_{k=1}^{LR^{2}}\nu_{k}^{(g)} + a_{1}^{(g)}\\
d_{1}^{(g)} &=&\frac{S_{g}}{2}\sum_{k=1}^{LR^{2}}\nu_{k}^{(g)}(\mu_{k}^{2(g)}+s_{k}^{2(g)})+
\frac{1}{2}\sum_{s:\eta_{s}=g}\Tr\left(diag(\bf{\underline{\pi}}^{(g)})(U^{T(s)}U^{(s)}+\Sigma_{g}^{(s)})\right)\\
&&-\left(\underline{\pi}^{(g)}\circ (\sum_{s:\eta_{s}:=g}U_{g}^{(s)T})\right)\underline{\mu}^{(g)T}+b_{1}^{(g)}\\
c_{0}^{(g)} &=& \frac{S_{g}}{2}\sum_{k=1}^{LR^{2}}(1-\nu_{k}^{(g)}) + a_{0}^{(g)}\\
d_{0}^{(g)} &=&\frac{S_{g}}{2}\sum_{k=1}^{LR^{2}}(1-\nu_{k}^{(g)})(\mu_{k}^{2(g)}+s_{k}^{2(g)})+
\frac{1}{2}\sum_{s:\eta_{s}=g}\Tr\left(diag(1-\bf{\underline{\pi}}^{(g)})(U^{T(s)}U^{(s)}+\Sigma_{g}^{(s)})\right)\\
&&-\left((1-\underline{\pi}^{(g)})\circ (\sum_{s:\eta_{s}:=g}U_{g}^{(s)T})\right)\underline{\mu}^{(g)T}+b_{0}^{(g)}
				,
\end{eqnarray*}
where $\underline{\pi}^{(g)}=[\pi_{1}^{(g)},\ldots,\pi_{LR^{2}}^{(g)}]$ and $\underline{\mu}^{(g)}=[\mu_{1}^{(g)},\ldots,\mu_{LR^{2}}^{(g)}]$ are $1\times LR^{2}$ vectors, and $S_{g}$ denotes the number of subjects belonging to group $g$.
\item \textbf{Update $\{\mu_{k}^{(g)},s_{k}^{2(g)}\}$ of $\tilde{\omega}_{k}^{(g)}$ and $\nu_{k}^{(g)}$ of $\gamma_{k}^{(g)}$ for $k=1,\ldots LR^{2}, g=1,\ldots, G$}:
For each group $g$, $\{\mu_{k}^{(g)},s_{k}^{2(g)}\}$ is updated as
\begin{eqnarray*}
\mu_{k}^{(g)} &=& \left(S_{g}\frac{c_{1}^{(g)}}{d_{1}^{(g)}}+\frac{\sum_{k'=1}^{LR^{2}}S_{kk'}}{q}\right)^{-1}\\
s_{k}^{2(g)} &=& \left(S_{g}\frac{c_{1}^{(g)}}{d_{1}^{(g)}}+\frac{\sum_{k'=1}^{LR^{2}}S_{kk'}}{q}\right)^{-1}\left(\frac{c_{1}^{(g)}}{d_{1}^{(g)}}\sum_{s:\eta_{s}=g}U_{g}^{(s)}(k)+
\frac{\sum_{k'=1}^{LR^{2}}S_{kk'}\pi_{k'}^{(g)}\mu_{k'}^{(g)}}{q}\right).
\end{eqnarray*}
By performing the necessary marginalization procedures, we have that the log odds $\rho_{k}^{(g)}$  is equal to
\begin{eqnarray*}
\rho_{k}^{(g)}& =&-\frac{S_{g}}{2}\left(\ln(d_{1}^{(g)})-\ln(d_{0}^{(g)})-\psi(c_{1}^{(g)})+\psi(c_{0}^{(g)})\right)\\
& &+\frac{\left(\sum_{s:\eta_{g}=g}U_{g}^{(s)}(k)\right)^{2}+\sum_{s:\eta_{g}=g}\Sigma_{g}^{(s)}(k)}{2}\left(\frac{c_{0}^{(g)}}{d_{1}^{(g)}}-\frac{c_{1}^{(g)}}{d_{1}^{(g)}}\right)\\
& &+\alpha_{0}^{(g)}+\mu_{\alpha_{1}^{(g)}}N_{k}^{(g)}  -\frac{\left(\sum_{k'=1}^{LR^{2}}S_{kk'}\pi_{k'}\mu_{k'}^{(g)}\right)^{2}+\sum_{k'=1}^{LR^{2}}S_{kk'}\sigma_{\alpha_{1}^{(g)}}^{2}}{2q\sum_{k'=1}^{LR^{2}}S_{kk'}} \\
& &-\frac{1}{2}\log\left(\frac{qS_{g}+\mathbb{E}[\xi_{1}^{(g)}]\sum_{k'=1}^{LR^{2}}S_{kk'}}{\mathbb{E}[\xi_{1}^{(g)}]\sum_{k'=1}^{LR^{2}}S_{kk'}}\right)+\frac{q\left(\left(\sum_{s:\eta_{g}=g}U_{g}^{(s)}(k)\right)^{2}+\sum_{s:\eta_{g}=g}\Sigma_{g}^{(s)}(k)\right)}{2\mathbb{E}[\xi_{1}^{(g)}]\left(qS_{g}+\mathbb{E}[\xi_{1}^{(g)}]\sum_{k'=1}^{LR^{2}}S_{kk'}\right)} \\
& &+\frac{2\left(\sum_{s:\eta_{g}=g}U_{g}^{(s)}(k)\right)\mathbb{E}[\xi_{1}^{(g)}]\left(\sum_{k'=1}^{LR^{2}}S_{kk'}\pi_{k'}\mu_{k'}^{(g)}\right)}{2\mathbb{E}[\xi_{1}^{(g)}]\left(qS_{g}+\mathbb{E}[\xi_{1}^{(g)}]\sum_{k'=1}^{LR^{2}}S_{kk'}\right)} \\
& &+\frac{\mathbb{E}[\xi_{1}^{2(g)}]\left(\left(\sum_{k'=1}^{LR^{2}}S_{kk'}\pi_{k'}\mu_{k'}^{(g)}\right)^{2}+\sum_{k'=1}^{LR^{2}}S_{kk'}\sigma_{\alpha_{1}^{(g)}}^{2}\right)}{2q\mathbb{E}[\xi_{1}^{(g)}]\left(qS_{g}+\mathbb{E}[\xi_{1}^{(g)}]\sum_{k'=1}^{LR^{2}}S_{kk'}\right)},
\end{eqnarray*}
where $\psi$ denotes the digamma function. It can be shown that the last three lines of the equation is an intractable function of $\xi_{1}^{(g)}$. As an approximation, we generate $1,000$ samples of  $\xi_{1}^{(g)}$ from $IG(c_{1}^{(g)},d_{1}^{(g)})$ and obtain a Monte Carlo estimate for $\rho_{k}^{(g)}$. If the model is fitted without external structural data, the terms $\alpha_{0}^{(g)}+\mu_{\alpha_{1}^{(g)}}N_{k}^{(g)}$ shown in the third line of the update rule for $\rho_{k}^{(g)}$ is replaced with $\psi(m^{(g)})-\psi(n^{(g)})$.

\item \textbf{Update $\{\mu_{\alpha_{1}^{(g)}},\sigma_{\alpha_{1}^{(g)}}^{2}\}$ of $\alpha_{1}^{(g)}$, for $g=1,\ldots G$}:
We follow the data augmentation procedure described in \eqref{eqn:13} to obtain closed from updates for $\{\mu_{\alpha_{1}^{(g)}},\sigma_{\alpha_{1}^{(g)}}^{2}\}$. Update is performed one group at as
\begin{eqnarray*}
\mu_{\alpha_{1}^{(g)}} &=&\frac{\sum_{k=1}^{LR^{2}}\left(\left[\nu_{k}^{(g)}-\frac{1}{2}-\phi_{k}^{(g)}\alpha_{0}^{(g)}\right]N_{k}^{(g)}\right)+\frac{w^{(g)}}{\tau^{2(g)}}}{\sum_{k=1}^{LR^{2}}\phi_{k}^{(g)}N_{k}^{2(g)}+\frac{1}{\tau^{2(g)}}} \\
\sigma_{\alpha_{1}^{(g)}}^{2}&=&\left(\sum_{k=1}^{LR^{2}}\phi_{k}^{(g)}N_{k}^{2(g)}+\frac{1}{\tau^{2(g)}}\right)^{-1}.
\end{eqnarray*}

\item \textbf{Update $\phi_{k}^{(g)}$ for $k=1,\ldots, LR^{2}, g=1,\ldots, G$}: We follow the results from \cite{Polson13} and directly compute the expectation of $\phi_{k}^{(g)}$ as
\begin{equation*}
\mathbb{E}[\phi_{k}^{(g)}]=\frac{\tanh\left((\alpha_{0}^{(g)}+\mathbb{E}[\alpha_{1}^{(g)}]N_{k}^{(g)})/2\right)}{2(\alpha_{0}^{(g)}+\mathbb{E}[\alpha_{1}^{(g)}]N_{k}^{(g)})}.
\end{equation*}

\item \textbf{Update $\{m^{(g)},n^{(g)}\}$ of $\pi^{(g)}$, for $g=1,\ldots G$} (this step replaces steps 5 and 6 if no external structural data is provided): By exploiting conjugacy, we update the pairs as 
\begin{eqnarray*}
m^{(g)} &=&e^{(g)}+\sum_{k=1}^{LR^{2}}\nu_{k}^{(g)} \\
n^{(g)}&=&f^{(g)}+LR^{2}-\sum_{k=1}^{LR^{2}}\nu_{k}^{(g)}.
\end{eqnarray*}

\end{enumerate}

\end{appendix}

\end{document}